\documentclass[preprint,12pt]{elsarticle}

\usepackage{graphicx} 
 \usepackage{booktabs}
 \usepackage{tabularx}

\usepackage{amssymb}
\usepackage{amsmath}
 \usepackage{lineno}
\journal{Nuclear Instruments and Methods in Physics Research Section A}

\begin{document}

\begin{frontmatter}



\title{First overnight balloon flight of the GRAINE 2023 emulsion gamma-ray telescope enabled by a large-scale pressure-vessel gondola}


\author[imass,kmi,nagoya]{Hiroki Rokujo\corref{cor1}}
\author[kobe]{Shigeki Aoki}
\author[kobe]{Takashi Azuma}
\author[nagoya]{Hirotaka Hayashi}
\author[nagoya]{Yudai Isayama}
\author[okayama]{Atsushi Iyono}
\author[kobe]{Takumi Kato}
\author[nagoya]{Tsuyoshi Kawahara}
\author[aichi]{Kohichi Kodama}
\author[nagoya]{Ryosuke Komatani}
\author[nagoya]{Masahiro Komatsu}
\author[nagoya]{Masahiro Komiyama}
\author[nagoya]{Hideyuki Minami}
\author[nagoya]{Kunihiro Morishima}
\author[okayama]{Fumiya Murakami}
\author[nagoya]{Shogo Nagahara}
\author[nagoya]{Naotaka Naganawa}
\author[nagoya]{Mitsuhiro Nakamura}
\author[nagoya]{Tomoaki Nakamura}
\author[kmi,nagoya]{Yuya Nakamura}
\author[nagoya]{Noboru Nakano}
\author[nagoya,imass,kmi]{Toshiyuki Nakano}
\author[gifu,fukui]{Kazuma Nakazawa}
\author[kobe]{Miyuki Oda}
\author[kobe]{Kazuhiro Okamoto}
\author[nagoya,kmi]{Osamu Sato}
\author[nagoya]{Kai Shimizu}
\author[nagoya]{Amon Suganami}
\author[okayama]{Yuki Sugi}
\author[imass,nagoya]{Kou Sugimura}
%\author[kobe-sci]{Atsumu Suzuki}
\author[kobe]{Satoru Takahashi}
\author[nagoya]{Ikuya Usuda}
\author[imass,nagoya]{Saya Yamamoto}
\author[kobe]{Jun Yamashita}
\author[kobe]{Mayu Yamashita}
%\author[okayama]{Riku Yamazaki}
\author[kobe]{Shoma Yoneno}
\author[gifu,label3]{Masahiro Yoshimoto}
\fntext[label3]{present address: Nishina Center for Accelerator-Based Science, RIKEN, 2-1 Hirosawa, Wako, Saitama 351-0198, Japan}

\cortext[cor1]{rokujo@nagoya-u.jp}

\affiliation[imass]{
    organization={Institute of Materials and Systems for Sustainability (IMaSS), Nagoya University},
    addressline={Furo-cho, Chikusa-ku},
    city={Nagoya},
    postcode={464-8601},
    country={Japan}
}

\affiliation[kmi]{
    organization={Kobayashi-Maskawa Institute for the Origin of Particles and the Universe (KMI), Nagoya University},
    addressline={Furo-cho, Chikusa-ku},
    city={Nagoya},
    postcode={464-8602},
    country={Japan}
}

\affiliation[nagoya]{
    organization={Graduate School of Science, Nagoya University},
    addressline={Furo-cho, Chikusa-ku},
    city={Nagoya},
    postcode={464-8602},
    country={Japan}
}
\affiliation[kobe]{
    organization={Graduate School of Human Development and Environment, Kobe University},
    addressline={3-11 Tsurukabuto, Nada-ku},
    city={Kobe},
    postcode={657-8501},
    country={Japan}
}

\affiliation[okayama]{
    organization={Graduate School of Science and Engineering, Okayama University of Science},
    addressline={1-1 Ridaicho, Kita-ku},
    city={Okayama},
    postcode={700-0005},
    country={Japan}
}
\affiliation[aichi]{
    organization={Faculty of Education, Aichi University of Education},
    addressline={1 Hirosawa, Igaya-cho},
    city={Kariya},
    postcode={448-8542},
    country={Japan}
}
\affiliation[gifu]{
    organization={Faculty of Education, Gifu University},
    addressline={1-1 Yanagido},
    city={Gifu},
    postcode={501-1193},
    country={Japan}
}
\affiliation[fukui]{
    organization={The Research Institute of Nuclear Engineering (RINE), University of Fukui},
    addressline={1-3-33 Kanawa, Tsuruga},
    city={Fukui},
    postcode={914-0055},
    country={Japan}
}

%\affiliation[kobe-sci]{
%    organization={Graduate School of Science, Kobe University},
%    addressline={1-1 Rokkodai-cho, Nada-ku},
%    city={Kobe},
%    postcode={657-8501},
%    country={Japan}
%}

%% Abstract
\begin{abstract}
The Gamma-Ray Astro Imager with Nuclear Emulsion (GRAINE) project conducts precision observations of cosmic gamma rays in the sub-GeV--GeV energy range using a balloon-borne nuclear-emulsion gamma-ray telescope with high angular resolution. In GRAINE 2023, a telescope with a total aperture area of 2.5 m$^{2}$ was flown in the first overnight balloon flight of the project, including observation periods for the Vela pulsar and the Galactic center region. To operate this large-area telescope under the low-pressure and low-temperature conditions of the stratosphere, the balloon-style pressure-vessel concept was scaled up, and a lightweight pressure-vessel gondola with an internal length of 4.9 m was developed. A new aluminum-alloy gondola ring structure was mechanically validated, and a lightweight membranous-shell material, SHL-300MDL, was developed. While increasing the telescope aperture area by a factor of 6.6 compared with GRAINE 2018, the mass of the pressure-vessel gondola was limited to 179 kg.
Ground tests using the completed flight assembly demonstrated that the pressure vessel maintained a differential pressure above 100 hPa at room temperature and under low-temperature conditions down to a mean temperature of $-66.0^{\circ}$C. The GRAINE 2023 payload was launched from Alice Springs, Australia, in April 2023 and achieved a total flight duration of approximately 27 h, including 24.3 h of level flight. Although the upper membranous shell reached approximately $-60^{\circ}$C during the night, the absolute pressure inside the vessel remained above the required minimum of 100 hPa throughout the level-flight period. These results demonstrate that the developed large and lightweight pressure-vessel gondola can accommodate a 2.5-m$^{2}$ emulsion gamma-ray telescope and maintain the required internal pressure under the low-temperature stratospheric conditions encountered during an overnight flight. Scientific analyses of astrophysical and atmospheric gamma rays, including a dedicated analysis of the Galactic center region, are ongoing using the recovered emulsion data. The present development provides a technical basis for repeated scientific observations with future large-area GRAINE telescopes.

\end{abstract}



%% Keywords

\end{frontmatter}



\section{Introduction}\label{sec1}
Cosmic gamma-ray observations in the sub-GeV--GeV energy range play an important role in understanding the acceleration and propagation of cosmic rays and high-energy astrophysical phenomena. The Fermi Gamma-ray Space Telescope, which has been in operation since 2008, has provided high-quality all-sky gamma-ray data in the 0.1–100 GeV energy range \cite{Fermi}. In the Galactic center region, however, an unexplained emission component known as the GeV excess has been reported even after subtracting contributions from known sources and diffuse emission \cite{Daylan}. Possible origins, including dark-matter annihilation and unresolved populations of millisecond pulsars, have been discussed \cite{Calore,Ackermann}, but its detailed spatial structure and origin remain uncertain because of limitations in angular resolution. Gamma-ray polarization in the sub-GeV--GeV range is also an important observable for probing emission mechanisms and magnetic-field structures in astrophysical sources \cite{Zhang}. However, polarization of astrophysical gamma rays has not yet been measured in this energy range, leaving it largely unexplored.
\par
The Gamma-Ray Astro Imager with Nuclear Emulsion (GRAINE) project aims at precise observations in the sub-GeV--GeV energy range using a balloon-borne gamma-ray telescope based on nuclear emulsion films \cite{TakahashiASR}. Nuclear emulsion is a tracking detector capable of recording charged-particle trajectories three-dimensionally with submicron spatial resolution \cite{Ariga}. By measuring the electron and positron tracks immediately downstream of the gamma-ray conversion point, the incident gamma-ray direction can be reconstructed with high precision, providing an angular resolution approximately one order of magnitude better than that of Fermi-LAT \cite{Nakamura}. In addition, the azimuthal angle of the pair-production plane defined by the electron and positron tracks can be reconstructed, allowing the polarization degree and polarization angle of incident gamma rays to be determined from the polarization-dependent modulation of its distribution \cite{Ozaki}. The development of the Hyper Track Selector (HTS), a high-speed automated track-readout system, has further enabled full-area readout and systematic analysis of large-area emulsion films \cite{Yoshimoto,Rokujo2018}. Using these technologies, the GRAINE 2018 balloon experiment conducted in Australia successfully imaged gamma rays from the Vela pulsar in the 0.1-GeV energy range with a 0.38-m$^2$-aperture emulsion telescope, demonstrating the capability of a balloon-borne emulsion telescope for astronomical observations \cite{TakahashiAPJ,Nakamura2021}.

\par

A GRAINE observational payload consists of an emulsion converter consisting of multiple converter units, each comprising a vacuum-packed stack of approximately 100 emulsion films \cite{Rokujo2018}; an emulsion multi-stage shifter that assigns timing information to individual events \cite{TakahashiNIM,Rokujo2012}; a star-camera-based attitude monitor; and a pressure-vessel gondola that enables these instruments to operate in the stratospheric environment \cite{Rokujo2019}. In each converter unit, incident gamma rays are converted into electron–positron pairs, whose tracks are recorded in the emulsion films. The multi-stage shifter consists of several movable stages, each carrying a vacuum-packed unit containing two emulsion films. By independently moving the stages, intentional positional offsets are introduced between corresponding tracks recorded in different stages, and the event time is reconstructed from the combination of these offsets. In both the converter and the multi-stage shifter, maintaining stable relative positions between the films within each vacuum pack is essential for reliable track connection across the emulsion layers. At ground level, atmospheric pressure compresses the vacuum packs and holds the stacked films together as a mechanically integrated unit. At balloon float altitude, however, the ambient atmospheric pressure decreases to only a few hPa. Therefore, to suppress relative displacement between the films and maintain stable track connection and angular measurement performance, the absolute pressure inside the pressure vessel must be maintained above 100 hPa during observations \cite{Rokujo2019}.
\par
Following the comprehensive demonstration of the telescope concept through the imaging of the Vela pulsar in GRAINE 2018, we pursued the development of a larger-aperture telescope and longer-duration balloon flights to obtain sufficient statistics for full-scale scientific observations. These developments have included a large-area, lightweight roller-driven multi-stage shifter \cite{Oda} and the establishment of mass-production techniques for large-area nuclear emulsion films \cite{Rokujo2024}. Building on these developments, the 2023 GRAINE balloon experiment (GRAINE 2023) employed two emulsion telescope units, each with an aperture area of 1.25 m$^2$. The total aperture area was thereby increased from 0.38 m$^2$ in GRAINE 2018 to 2.5 m$^2$, corresponding to an increase by a factor of 6.6. In addition to the Vela pulsar, the Galactic center region, which entered the telescope field of view from around midnight to early morning in late April at Alice Springs, Australia, was selected as an observation target. This required the first overnight flight of GRAINE, with launch in the early morning and continued flight until after sunrise on the following day.
\par
The GRAINE 2023 payload was launched from the Alice Springs Balloon Launching Station, Australia, on April 30, 2023, and successfully completed the first overnight flight of the GRAINE project, with a total flight duration of approximately 27 h. This paper reports the development, ground testing, and in-flight performance of the large and lightweight pressure-vessel gondola that enabled this flight. Section 2 describes the configuration of the GRAINE 2023 payload and pressure-vessel gondola, the development of the gondola ring structure and lightweight shell, and the mechanical and pressure-retention tests. Section 3 presents an overview of the balloon flight and the temperature and internal-pressure profiles during the overnight flight.

%%%%%%%%%%%%%%%%%%%%%%%%%%%%%%%%%%%%%%%%%%%%%%%%%%%
%%%%%%%%%%%%%%%%%%%%%%%%%%%%%%%%%%%%%%%%%%%%%%%%%%%

\section{Development and ground testing of the GRAINE 2023 pressure-vessel gondola}
\subsection{Payload configuration and balloon-style pressure-vessel concept}\label{subsec1}
\begin{figure}[t]%% placement specifier
%% Use \includegraphics command to insert graphic files. Place graphics files in 
%% working directory.
\centering%% For centre alignment of image.
\includegraphics[width=0.95\linewidth]{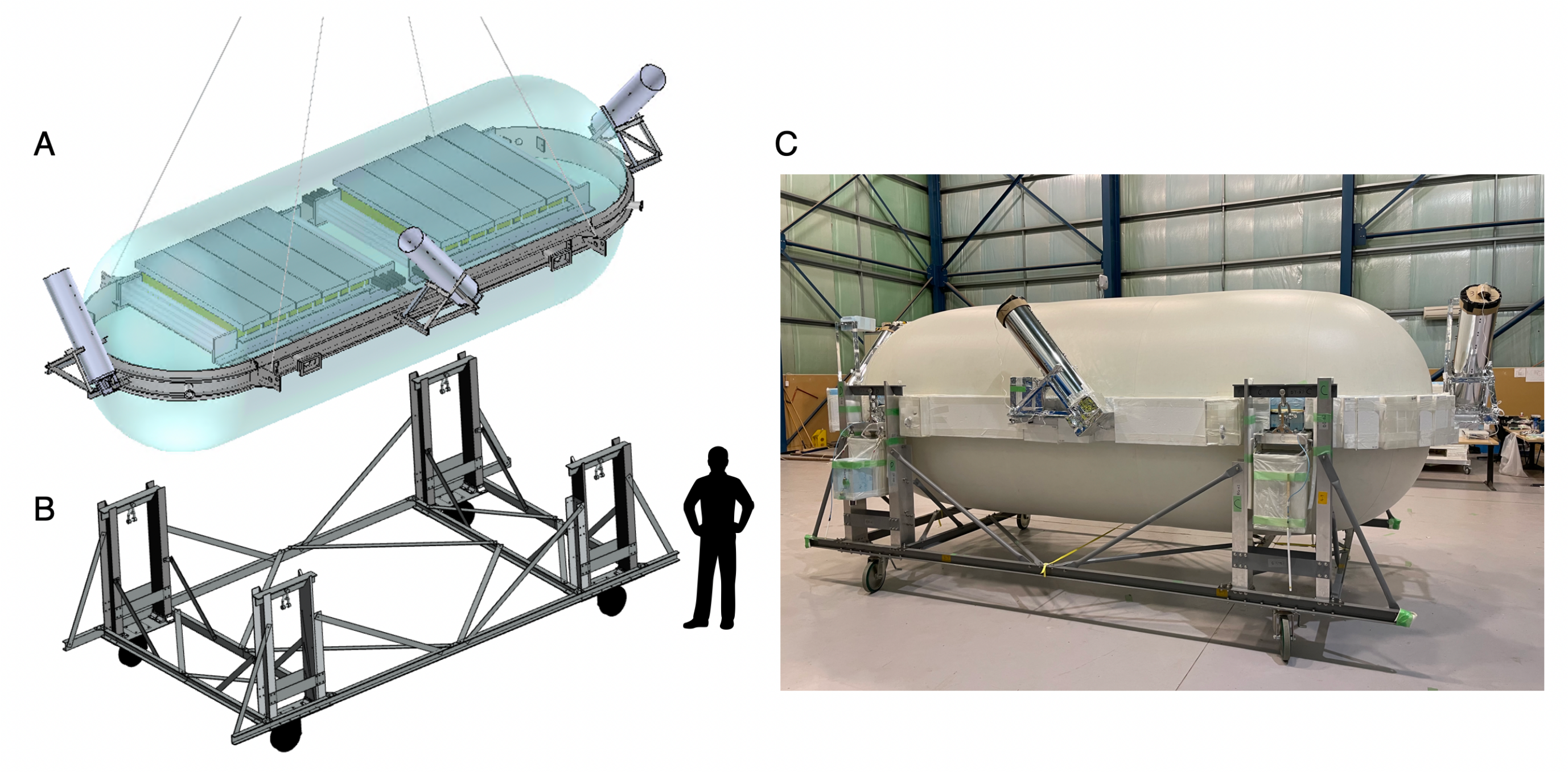}
%% Use \caption command for figure caption and label.
\caption{\label{fig:payload}Overview of the GRAINE 2023 pressure-vessel gondola and payload configuration.
(A) Three-dimensional view of the flight configuration, including two emulsion telescope units and three star-camera attitude-monitor units mounted on the pressure-vessel gondola.
(B) Transport cart used for ground handling and assembly. The cart was separated from the pressure-vessel gondola before flight.
(C) Photograph of the fully assembled GRAINE 2023 pressure-vessel gondola mounted on the transport cart during ground preparation.}
\end{figure}

Figure \ref{fig:payload} shows an overview of the GRAINE 2023 payload and pressure-vessel gondola. GRAINE 2023 carried two emulsion telescope units, each with an aperture area of 1.25 m$^2$, providing a total aperture area of 2.5 m$^2$. Figure \ref{fig:payload}(A) shows the payload configuration during flight. In addition to the two emulsion telescope units, the long-cylinder pressure-vessel gondola carried three star-camera-based attitude-monitor units, control electronics, and batteries. To accommodate these instruments, the pressure-vessel gondola was designed with an overall length of approximately 5 m.
\par
Figure \ref{fig:payload}(B) shows the dedicated transport cart used for ground transportation, assembly, and handling. In GRAINE 2018, the truss frame used to support and transport the gondola on the ground remained attached to the pressure-vessel gondola during flight \cite{Rokujo2019}. In GRAINE 2023, these ground-support functions were transferred to a separate transport cart to reduce the flight mass, and the cart was detached from the pressure-vessel gondola before launch. Figure \ref{fig:payload}(C) shows the fully assembled GRAINE 2023 pressure-vessel gondola mounted on the transport cart during ground preparation.

\begin{figure}[htbp]
\centering % \begin{center}/\end{center} takes some additional vertical space
\includegraphics[width=0.95\linewidth]{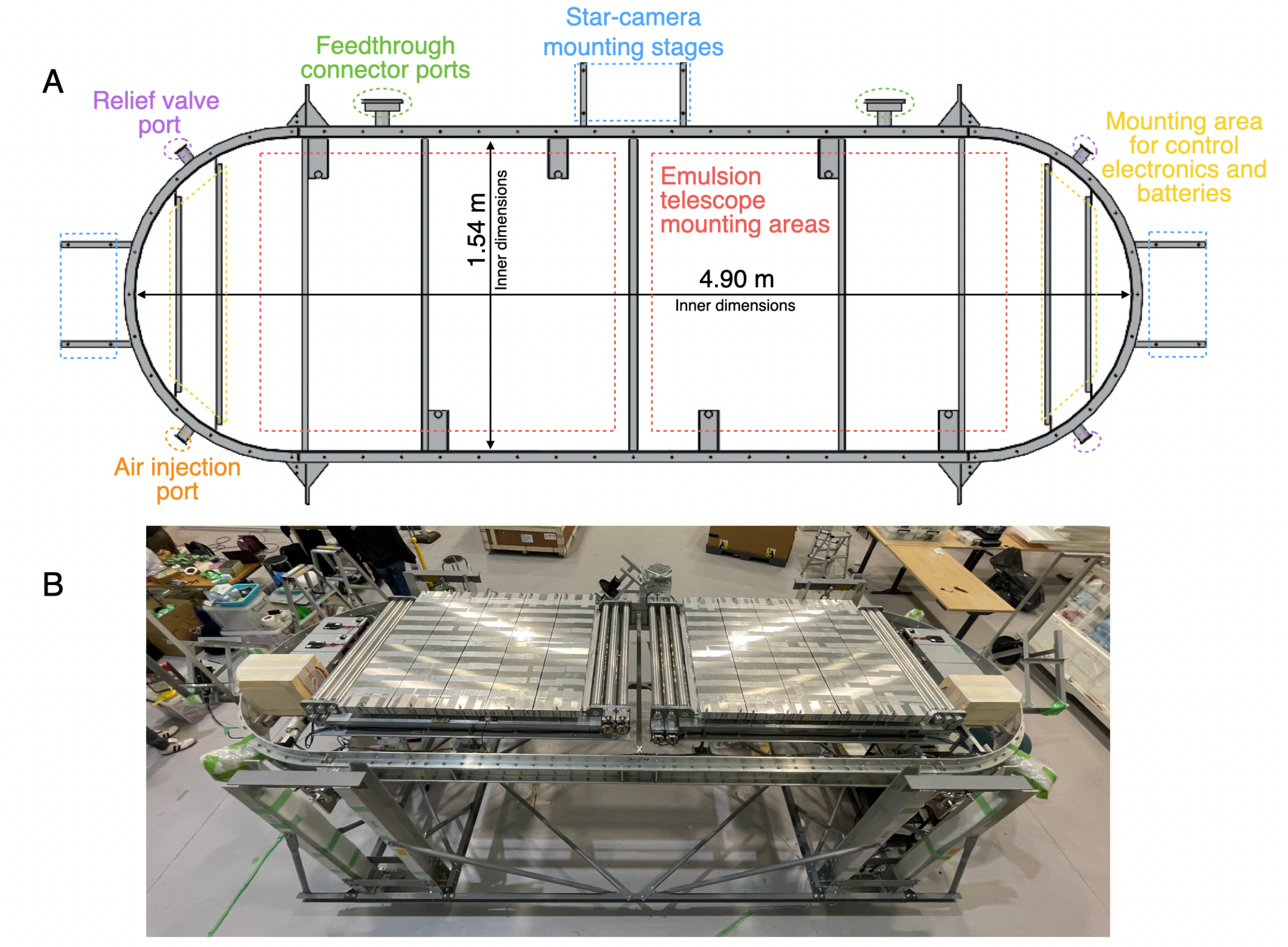}
% "\includegraphics" from the "graphicx" permits to crop (trim+clip)
% and rotate (angle) and image (and much more)
\caption{\label{fig:ring}Layout of major payload components on the gondola ring structure of the GRAINE 2023 pressure-vessel gondola.
(A) Schematic top view of the gondola ring structure, showing the mounting areas for the two emulsion telescope units, three star-camera attitude monitors, control electronics, and batteries, as well as the locations of the air-injection port, relief-valve port, and feedthrough connector ports. The inner dimensions of the main ring are 4.90 m $\times$ 1.54 m.
(B) Photograph of the two emulsion telescope units installed on the gondola ring structure before installation of the thermal insulation, airtight films, and membranous shells.}
\end{figure}

Figure \ref{fig:ring} shows the aluminum-alloy gondola ring structure, consisting of the main ring and sub-rings, together with the arrangement of the payload components. As shown in Figure \ref{fig:ring}(A), the gondola ring structure provided mounting areas for the two emulsion telescope units, three star-camera-based attitude monitors, control electronics, and batteries. It also incorporated an air-injection port, a port for the differential-pressure relief valve, and feedthrough connector ports for power and signal cables. The inner dimensions of the main ring were 4.90 m $\times$ 1.54 m. Figure \ref{fig:ring}(B) shows the two emulsion telescope units installed on the gondola ring structure before installation of the thermal insulation, airtight membrane, and membranous shell.

\begin{figure}[htbp]
\centering % \begin{center}/\end{center} takes some additional vertical space
\includegraphics[width=0.95\linewidth]{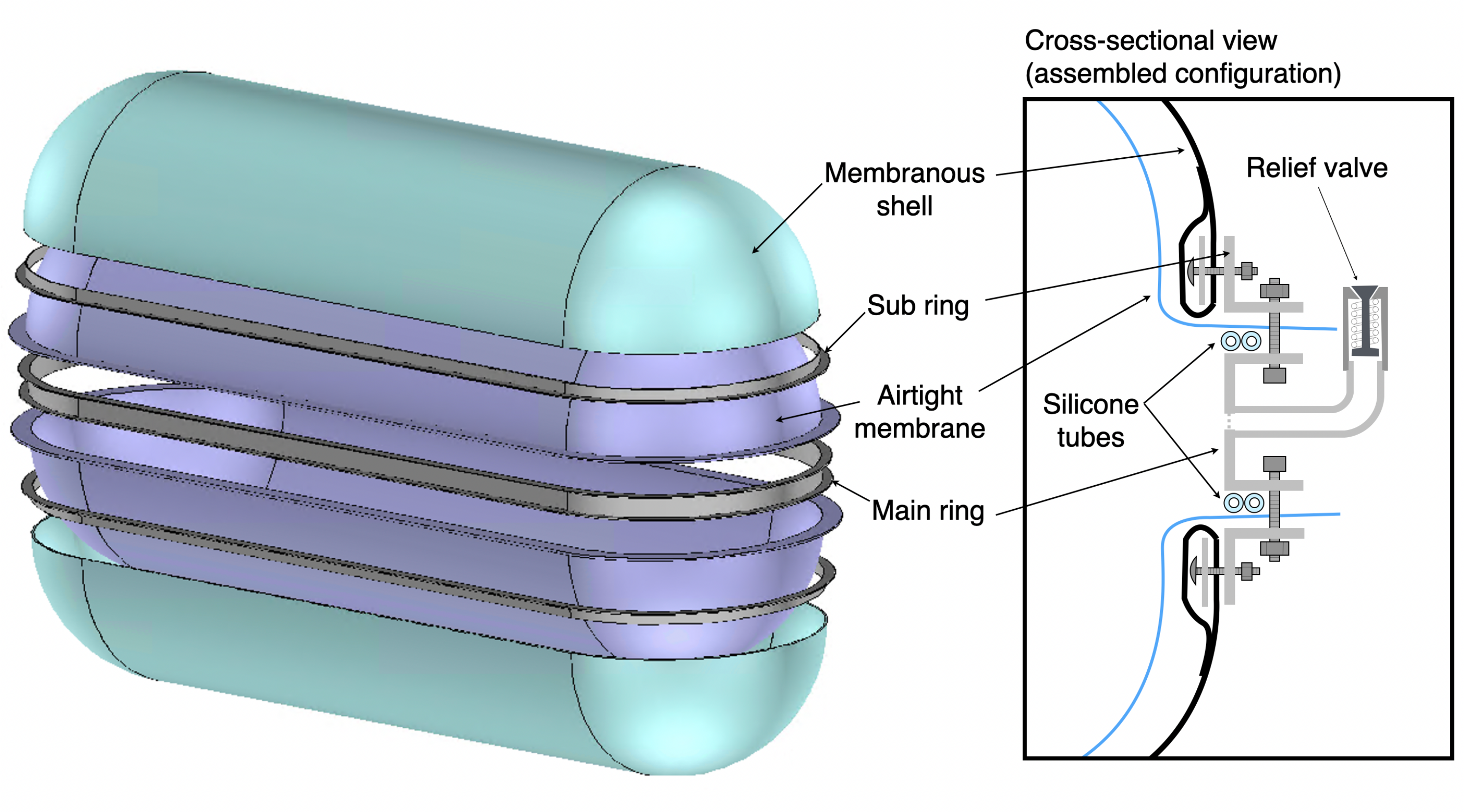}
% "\includegraphics" from the "graphicx" permits to crop (trim+clip)
% and rotate (angle) and image (and much more)
\caption{\label{fig:concept}Schematic illustration of the balloon-style long-cylinder pressure vessel adopted for GRAINE. The concept was originally proposed in Ref. \cite{Rokujo2019}. The left panel shows an exploded view of the vessel components, and the right panel shows a cross-sectional view of the sealing structure and relief valve in the assembled configuration.}
\end{figure}

GRAINE employs a balloon-style pressure-vessel concept to provide a stable operating environment for the emulsion telescope at balloon float altitude \cite{Rokujo2019}. Figure \ref{fig:concept} illustrates the concept. The pressure vessel consists of a main ring that supports the payload instruments, an airtight membrane that provides an airtight enclosure, a membranous shell that supports the load generated by the internal differential pressure through membrane tension, sub-rings that secure these membranes, and a differential-pressure relief valve that prevents excessive differential pressure during ascent. The airtight membrane is clamped between the main ring and the sub-rings, and airtightness is maintained using sealing elements and mechanical fasteners.
\par
In this design, the internal differential pressure is supported by flexible membrane materials rather than by a rigid metallic vessel. This provides the low mass required for a balloon-borne instrument and also allows the membrane structure to be transported in a compact form. In addition, the amount of material in the telescope field of view can be minimized, thereby reducing both absorption of incident gamma rays and the production of secondary particles through cosmic-ray interactions.
\par
The long-cylinder configuration allows the available instrument area to be increased by extending the straight section without substantially increasing the radius of curvature of the vessel cross section. A larger-area emulsion telescope can therefore be accommodated without the increase in membrane tension that would result from increasing the vessel diameter. This concept was demonstrated with a short-cylinder pressure vessel in GRAINE 2018 \cite{Rokujo2019}. In the present work, the concept was scaled up to the size required for the GRAINE 2023 scientific payload and implemented as a pressure-vessel gondola approximately 5 m in overall length.

%% Use \subsubsection, \paragraph, \subparagraph commands to 
%% start 3rd, 4th and 5th level sections.
%% Refer following link for more details.
%% https://en.wikibooks.org/wiki/LaTeX/Document_Structure#Sectioning_commands

\subsection{Design constraints and scale-up from GRAINE 2018}
The design of the GRAINE 2023 pressure-vessel gondola had to simultaneously satisfy the operating requirements of the emulsion telescope, the environmental conditions expected during an overnight flight, and the mass constraints of a balloon payload.
\par
As described in Section 1, the stacked emulsion films in the telescope are vacuum-packed and must be externally pressurized to suppress relative displacement between the films and thereby maintain reliable track connection and high-precision angular measurements. At balloon float altitude, the ambient atmospheric pressure decreases to only a few hPa. Based on the results of the previous study \cite{Rokujo2019}, the absolute pressure inside the pressure vessel was therefore required to remain above 100 hPa during flight.
\par
For ground qualification of the long-term pressure-retention capability, the vessel was required to maintain a differential pressure of at least 100 hPa for 40 h under conditions in which the external pressure remained nearly constant. Thus, whereas the in-flight requirement was specified in terms of the absolute pressure inside the vessel, the ground-test criterion was defined in terms of the differential pressure between the inside and outside of the vessel.
\par
During an overnight flight, the temperature of the pressure-vessel shell was expected to decrease to approximately $-60^{\circ}$C at night. The membranous shell, airtight membrane, heat-sealed joints, and sealing components were therefore required to retain sufficient flexibility and airtightness at low temperature to maintain the required internal pressure.
\par

GRAINE 2023 required a substantial increase in telescope aperture compared with GRAINE 2018. Simply scaling up the previous gondola design would have resulted in an excessive increase in structural mass. The basic design strategy was therefore to minimize the mass of each gondola component while maintaining the required mechanical strength, pressure-retention performance, and payload-mounting capability.
\par
The principal design changes introduced for mass reduction were the replacement of the flight-mounted ground-support and transport truss by the detachable transport cart described in Section 2.1, the development of a new aluminum-alloy gondola ring structure supporting the two telescope units, and the development of a lightweight membranous-shell material. Table  \ref{tab:comparison} compares the key specifications and mass breakdowns of the GRAINE 2018 and GRAINE 2023 payloads and pressure-vessel gondolas. 
\par
Although the total telescope aperture area was increased by a factor of 6.6, from 0.38 m$^2$ to 2.5 m$^2$, the mass of the pressure-vessel gondola increased only from 147 kg to 179 kg, corresponding to an increase of approximately 22\%. These developments enabled the 2.5-m$^2$ telescope to be integrated into a 703-kg GRAINE payload, consistent with the approximately 700 kg payload-mass target for the JAXA balloon experiment. The structural and material developments underlying this scale-up are described in the following sections.

\begin{table}[t]
\centering
\caption{\label{tab:comparison}Comparison of the key specifications and mass budgets of the
GRAINE 2018 and GRAINE 2023 payloads and pressure-vessel gondolas.}
\begin{tabular}{lcc}
\toprule
Parameter & GRAINE 2018 & GRAINE 2023 \\
\midrule

\multicolumn{3}{l}{\textit{Key specifications}} \\
Total telescope aperture area
    & 0.38~m$^{2}$ & 2.5~m$^{2}$ \\
Pressure-vessel shape
    & Short cylinder & Long cylinder \\
Internal vessel length
    & 2.1~m & 4.9~m \\
Membranous-shell material
    & SHL-300 & SHL-300MDL \\
Shell areal density
    & 0.1~g\,cm$^{-2}$ & 0.067~g\,cm$^{-2}$ \\

\midrule
\multicolumn{3}{l}{\textit{Mass budget}} \\
\textbf{Pressure-vessel gondola mass}
    & \textbf{147~kg} & \textbf{179~kg} \\
\hspace{1em}-- Aluminum ring structure
    & 75~kg & 123~kg \\
\hspace{1em}-- Ground-support and transport truss frame
    & 33~kg & Not used \\
\hspace{1em}-- Membranous shells and airtight membranes
    & 10~kg & 28~kg \\
\hspace{1em}-- Other pressure-vessel hardware$^{a}$
    & 29~kg & 28~kg \\
\textbf{Integrated GRAINE payload mass}
    & \textbf{348~kg} & \textbf{703~kg} \\

\bottomrule
\end{tabular}

\vspace{1mm}
\begin{flushleft}
\footnotesize
$^{a}$Includes battery mounting structures, shackles,
pressure-control valves and associated tubing,
shell-to-ring clamping hardware, fasteners,
and other associated flight hardware.
\end{flushleft}
\end{table}

\subsection{Lightweight design and structural evaluation of the gondola ring structure}

The GRAINE 2023 pressure-vessel gondola, approximately 5 m in overall length, was required to support two emulsion telescope units, each weighing approximately 200 kg, together with the associated payload components, while maintaining sufficient mechanical strength with minimum structural mass. In addition to the loads encountered during normal flight, resistance to transient loads associated with events such as parachute opening at flight termination was an important design consideration. The GRAINE 2023 gondola was designed with reference to the NASA/CSBF guideline for balloon gondola structures, OM-220-10-H \cite{CSBF}. One of the structural criteria specified in this guideline requires the load-bearing structure to withstand a vertical load equal to 10 times the payload weight without ultimate structural failure.
\par
Commercially available aluminum-alloy sections were used for the main ring, sub-rings, and internal crossbeams, which were fabricated by bending and welding. To reduce the structural mass, portions of the standard sections were removed longitudinally while retaining the required bending stiffness and mechanical strength. For example, a 125 $\times$ 65 $\times$ 6 mm C-channel was used for the main ring, and its flange width was reduced from 65 to 55 mm, resulting in a cross section of 125 $\times$ 55 $\times$ 6 mm. The completed gondola ring structure had a mass of 123 kg.
\par
The two emulsion telescope units were each supported at three points on the main ring. Four suspension points, fabricated from 15-mm-thick aluminum-alloy plates, were welded to the outer side of the main ring and used to suspend the entire payload. The three-point support of each telescope unit avoided mechanical overconstraint and reduced the transfer of local loads to the telescope structures due to deformation and torsion of the main ring under the four-point suspension. Changes in the orientation of each telescope unit before and after suspension were measured using inclinometers mounted on the units and were used to correct the relative orientation between the telescopes and the star-camera-based attitude monitors.
\par

To achieve sufficient mechanical strength under load conditions of approximately 10 times the payload weight while minimizing the structural mass, linear static finite-element analyses were repeatedly performed during the design process using SOLIDWORKS Simulation. The positions, number, and cross-sectional geometries of the main ring and internal crossbeams were examined, with the cross section of each crossbeam selected according to the load conditions at its location. The analyses showed that five internal crossbeams at the adopted positions were required to keep the stresses within the material-strength criterion under the tenfold-load condition, and this ring structure was adopted for GRAINE 2023.
\par
\begin{figure}[htbp]
\centering % \begin{center}/\end{center} takes some additional vertical space
\includegraphics[width=0.95\linewidth]{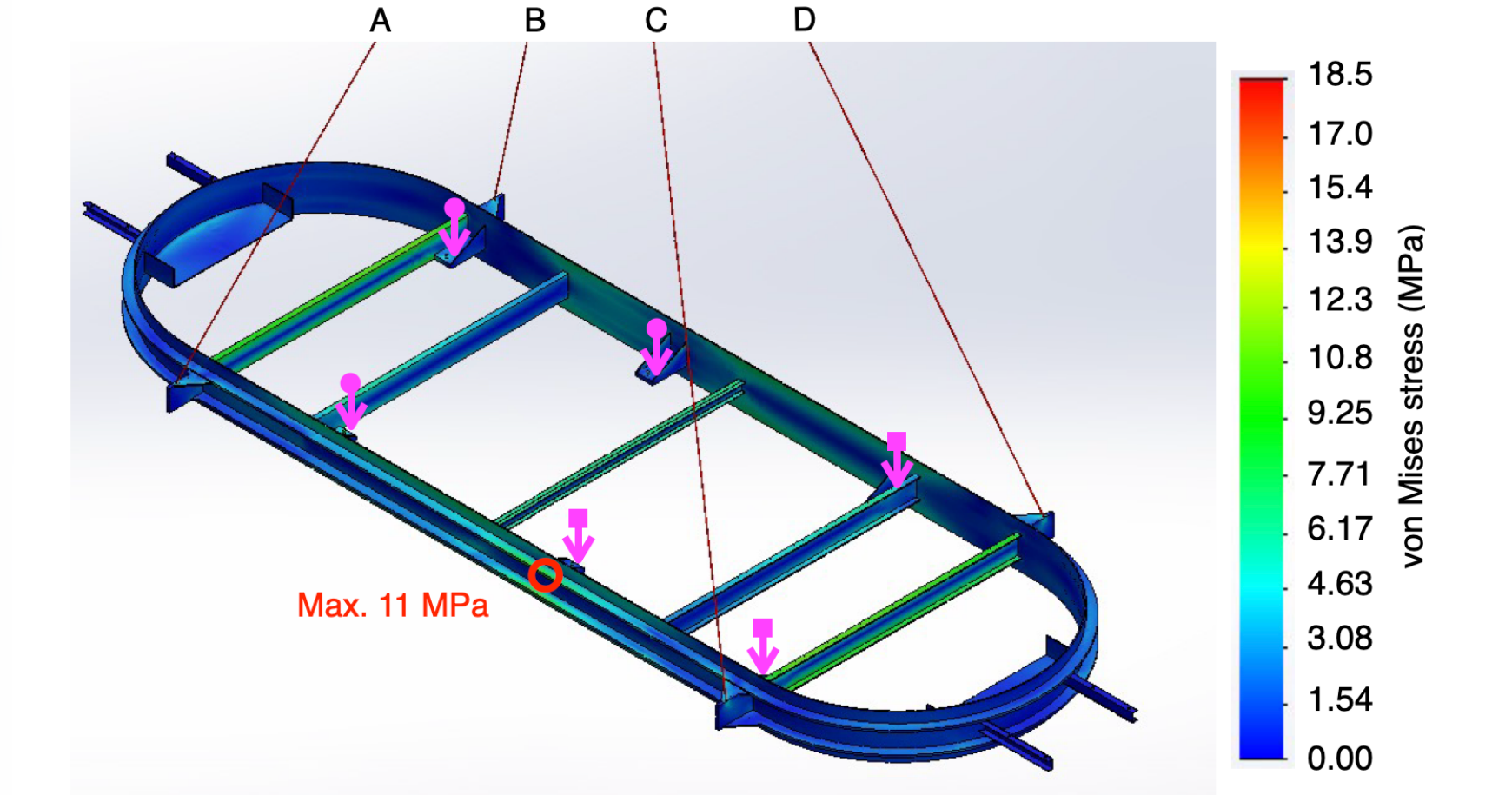}
% "\includegraphics" from the "graphicx" permits to crop (trim+clip)
% and rotate (angle) and image (and much more)

\caption{\label{fig:SOLIDWORKS}Finite-element model and von Mises stress distribution of the GRAINE 2023 gondola ring structure under the 1-G loading condition. The ring structure incorporates five internal crossbeams. Suspension rods A–D reproduce the four-point suspension configuration. Loads corresponding to the two emulsion telescope units were applied at their respective three-point mounting positions, as indicated by the magenta arrows. The red circle indicates the location of the maximum von Mises stress, approximately 11 MPa.
}
\end{figure}

Figure  \ref{fig:SOLIDWORKS} shows the finite-element analysis of the gondola ring structure adopted for GRAINE 2023. The structural material was modeled as A6063-T5 aluminum alloy. In addition to the self-weight of the gondola ring structure, loads representing the two emulsion telescope units and the other payload components expected in the flight configuration were included in the analysis. The loads of the two telescope units were applied at their respective three-point mounting positions. Rods A–D, representing the flight suspension slings, were connected to the four suspension points and constrained at a common point above the structure to reproduce the four-point suspension configuration. Figure  \ref{fig:SOLIDWORKS} shows the von Mises stress distribution under the 1-G loading condition. Because the stress scales linearly with the applied load in the linear static analysis, a vertical load equal to 10 times the payload weight was used as the design reference. Accordingly, one-tenth of the tensile strength of A6063-T5, 185 MPa, corresponding to 18.5 MPa, was used as the design criterion for the 1-G analysis. The maximum von Mises stress in the adopted ring structure was approximately 11 MPa, below this criterion. Linear scaling of the 1-G result to the tenfold-load condition gives a maximum stress of approximately 110 MPa, which is also below the 0.2\% proof stress of A6063-T5, 145 MPa. The analysis therefore indicated that the ring structure remained within the elastic range under the tenfold-load condition.
\par

\par

\par
\begin{figure}[htbp]
\centering % \begin{center}/\end{center} takes some additional vertical space
\includegraphics[width=0.95\linewidth]{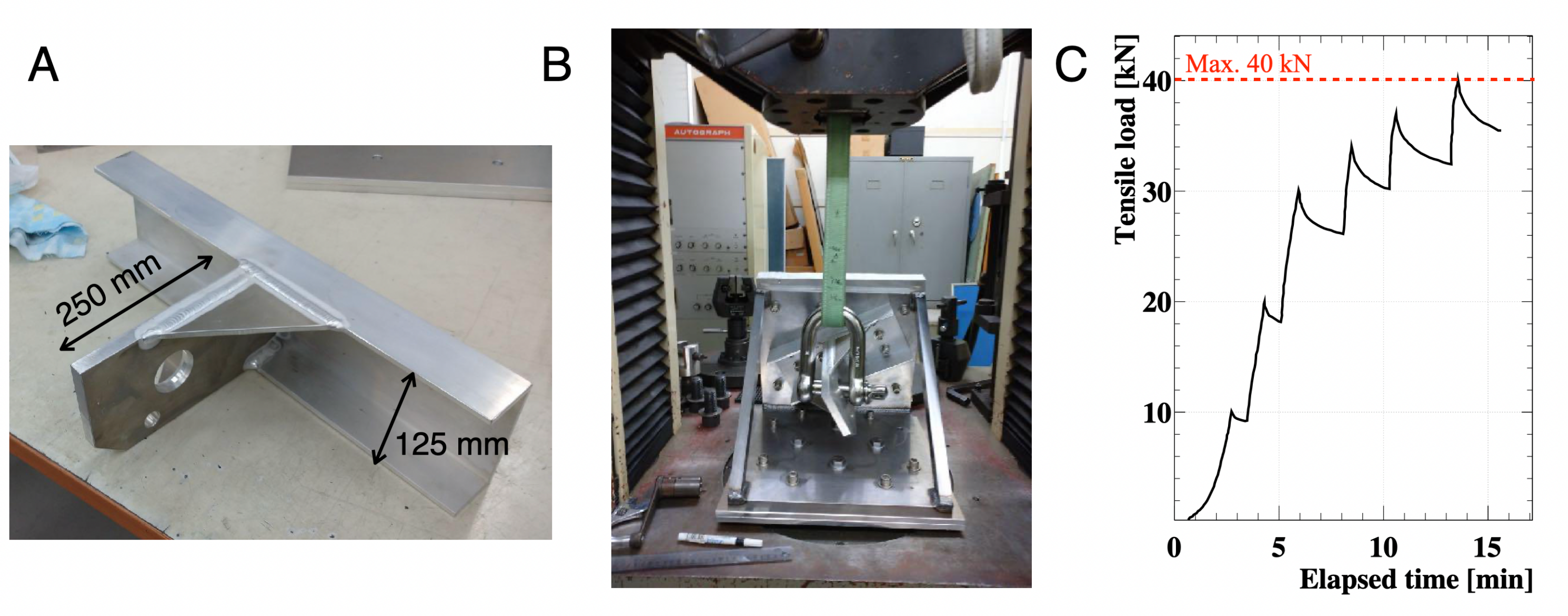}
% "\includegraphics" from the "graphicx" permits to crop (trim+clip)
% and rotate (angle) and image (and much more)
\caption{Full-scale tensile test of the welded suspension-point assembly. (A) Full-scale specimen fabricated using the same dimensions, materials, and welding procedure as the flight structure. (B) Tensile-test setup reproducing the suspension direction of the flight configuration. (C) Applied tensile-load history. The load was increased stepwise to 40 kN at a crosshead speed of 10 mm/min. No fracture or visible damage was observed in either the welded joints or the base material after the test.\label{fig:HangingPoint}
}
\end{figure}

For the suspension points welded to the main ring, the strength of the welded structure was further evaluated by a full-scale tensile test using a specimen fabricated to the same specifications as the flight suspension structure. Figures \ref{fig:HangingPoint}(A) and 5(B) show the test specimen and tensile-test setup, respectively, and Figure \ref{fig:HangingPoint}(C) shows the applied load history. For a reference payload mass of 800 kg, the tenfold vertical load specified in the NASA/CSBF guideline corresponds to a sling-direction load of approximately 21.5 kN at each suspension point when the flight suspension geometry is taken into account. In the test, the load was increased stepwise in the actual suspension direction to a maximum of 40 kN. The welded joint and base material were inspected at each loading step, and no fracture or visible damage was observed up to the maximum load. The maximum test load was approximately 1.9 times the load at each suspension point corresponding to the tenfold vertical-load condition. This test confirmed the mechanical integrity of the suspension-point geometry and welding procedure used for the flight structure.

\subsection{Development of the lightweight membranous shell}

In the long-cylinder pressure vessel, the differential pressure between the inside and outside of the vessel generates tensile force in the membranous shell. The circumferential tensile force per unit length, $\gamma$, in the cylindrical section is given by
\begin{equation}
 \gamma=\Delta P \times R
\end{equation}
where $\Delta P$ is the differential pressure and $R$ is the radius of curvature of the shell \cite{Rokujo2019}. For the GRAINE 2023 pressure-vessel gondola, the design differential pressure was 300 hPa and the radius of curvature of the cylindrical section was 0.77 m, giving a circumferential tensile force of 231 N/cm. Applying a safety factor of 2.5, the required strength of the membranous shell and its heat-sealed joints was set to 578 N/cm, corresponding to 1733 N for a 3-cm-wide specimen.
\par
Because the overall length of the GRAINE 2023 pressure vessel was increased to approximately 5 m, the area of the membranous shell was substantially larger than that of the previous gondola. To limit the associated increase in shell mass, a new lightweight composite membrane material, SHL-300MDL, was developed based on SHL-300 used in GRAINE 2018, with a reduced thickness of the polyurethane coating layer.
\par
\begin{figure}[htbp]
\centering % \begin{center}/\end{center} takes some additional vertical space
\includegraphics[width=0.95\linewidth]{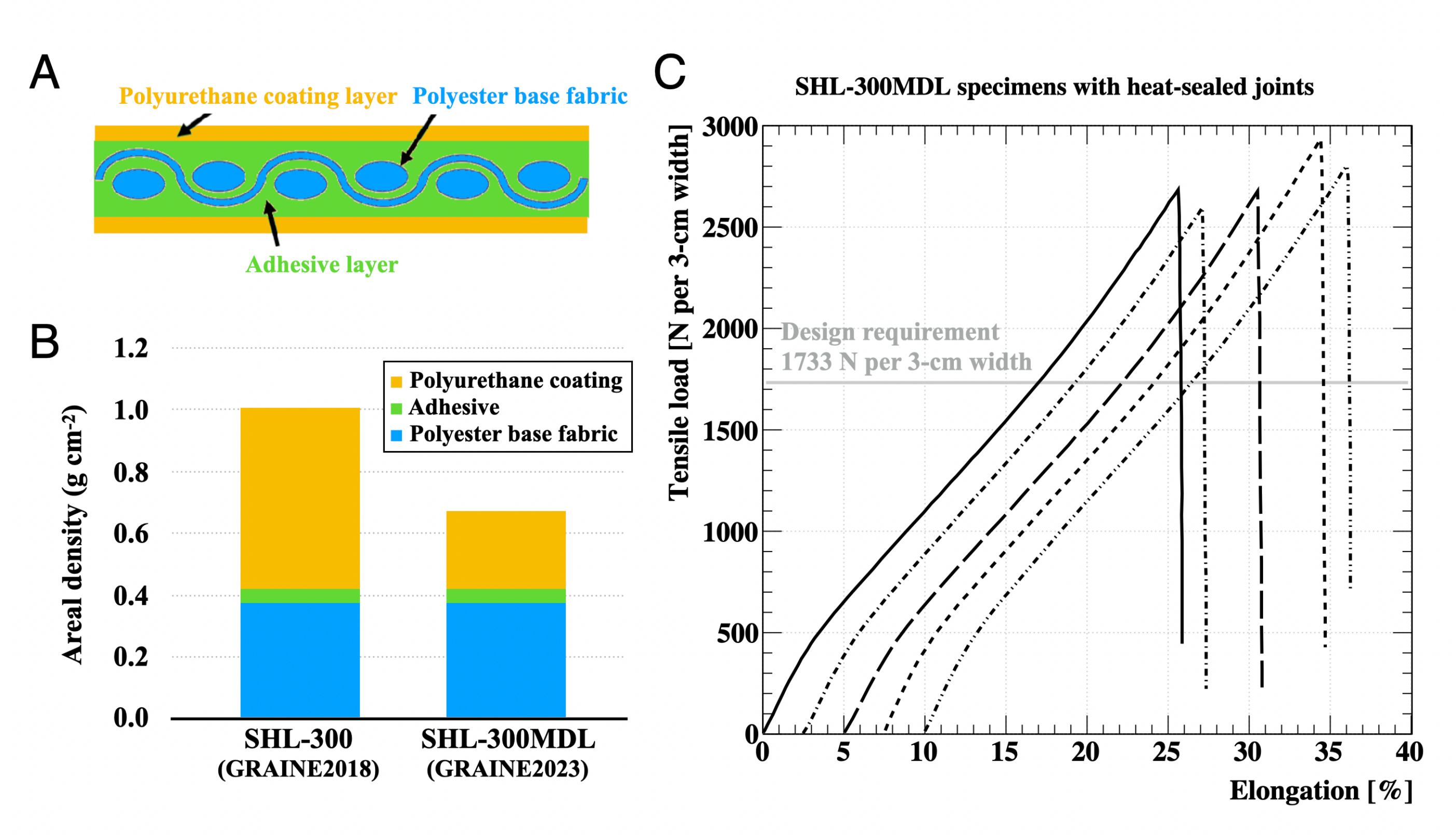}
% "\includegraphics" from the "graphicx" permits to crop (trim+clip)
% and rotate (angle) and image (and much more)
\caption{\label{fig:shell} Structure, areal density, and tensile performance of the shell membrane materials.
(A) Schematic cross-sectional structure of the composite shell material, consisting of a polyester base fabric, an adhesive layer, and polyurethane coating layers.
(B) Comparison of the areal densities of SHL-300, used for GRAINE 2018, and the lightweight SHL-300MDL developed for GRAINE 2023. The areal density was reduced from 1000 to 600 g/m$^{2}$, mainly by reducing the amount of polyurethane coating.
(C) Tensile load–elongation curves of five 3-cm-wide SHL-300MDL specimens containing a heat-sealed joint, measured at room temperature. The five curves are horizontally offset for clarity. All tested specimens exceeded the design requirement of 1733 N per 3-cm width before rupture.
}
\end{figure}

Figure \ref{fig:shell}(A) shows the cross-sectional structure of SHL-300MDL. The material consists of a polyester base fabric, adhesive layers, and polyurethane coating layers. By reducing the amount of polyurethane coating while retaining the same base-fabric structure, the areal density was reduced by 33\%, from 0.1 g/cm$^2$ for SHL-300 to 0.067 g/cm$^2$ for SHL-300MDL, as shown in Figure \ref{fig:shell}(B).
\par
To verify that SHL-300MDL provided sufficient strength for the membranous shell, tensile tests to rupture were performed on specimens containing heat-sealed joints. Each specimen was 3 cm wide, and the heat-sealed joints were fabricated under the same conditions used for the flight shell. Figure \ref{fig:shell}(C) shows the tensile load–elongation curves measured at room temperature for five specimens. The rupture loads ranged from 2604 to 2948 N per 3-cm width, with a mean value of 2748 N per 3-cm width. Although this mean rupture load was lower than the 3467 N per 3-cm width obtained for SHL-300\cite{Rokujo2019}, it remained well above the design requirement of 1733 N per 3-cm width.
\par
Constant-load tests were also performed on the heat-sealed joints to evaluate their performance under the low-temperature environment expected during nighttime flight and the high-temperature conditions that could occur during ground preparation and testing. A load of 1800 N per 3-cm width, corresponding approximately to the design requirement, was maintained for 1 h at both $-70^{\circ}$C and $+50^{\circ}$C. No peeling or rupture of the heat-sealed joints was observed under either condition. These tests demonstrated that SHL-300MDL satisfied the strength requirement for the GRAINE 2023 membranous shell while reducing the areal density by 40\%.

\subsection{Pressure retention performance under low-temperature conditions}

In this section, the long-term pressure-retention performance of the GRAINE 2023 pressure-vessel gondola was evaluated at room temperature, together with its pressure-retention performance under low-temperature conditions representative of an overnight flight.
\par
In GRAINE 2018, the airtight membrane was fabricated by cutting a natural-rubber meteorological balloon into a hemispherical shape. For GRAINE 2023, a new natural-rubber airtight membrane was fabricated to fit the approximately 5-m-long cylindrical vessel geometry. The sealing system employed tubular gaskets between the main and sub-rings and O-rings for the various ports, both made of Zymer SF-5, a low-temperature rubber material developed and evaluated in the previous study \cite{Rokujo2019}. Pressure-retention tests were performed at room and low temperatures using the flight pressure-vessel assembly, consisting of the airtight membrane, SHL-300MDL membranous shell, main ring, sub-rings, and sealing components. The differential-pressure relief valve was tested separately under low-temperature conditions.
\par
First, a long-duration pressure-retention test was performed at room temperature. The vessel was pressurized to a differential pressure of 310 hPa, after which the differential pressure, vessel temperature, and ambient temperature were continuously monitored. Although the differential pressure gradually decreased with time, it remained at 270 hPa after 40 h, well above the ground-test criterion of 100 hPa. The test was also repeated after separating the main and sub-rings, replacing the tubular gaskets, and reassembling the sealing structure, and comparable pressure-retention performance was reproduced.
\par
A full-scale low-temperature environmental test was then performed. Although component-level tests of the previous pressure-vessel model had indicated sufficient pressure-retention capability at approximately $-60 ^{\circ}$C, a low-temperature test using a completed pressure vessel approximately 5 m in length had not previously been conducted. Because the full-scale vessel could not be accommodated in the existing environmental chamber, a large refrigerated container was used to provide the required low-temperature environment.
\par
\begin{figure}[htbp]
\centering % \begin{center}/\end{center} takes some additional vertical space
\includegraphics[width=0.95\linewidth]{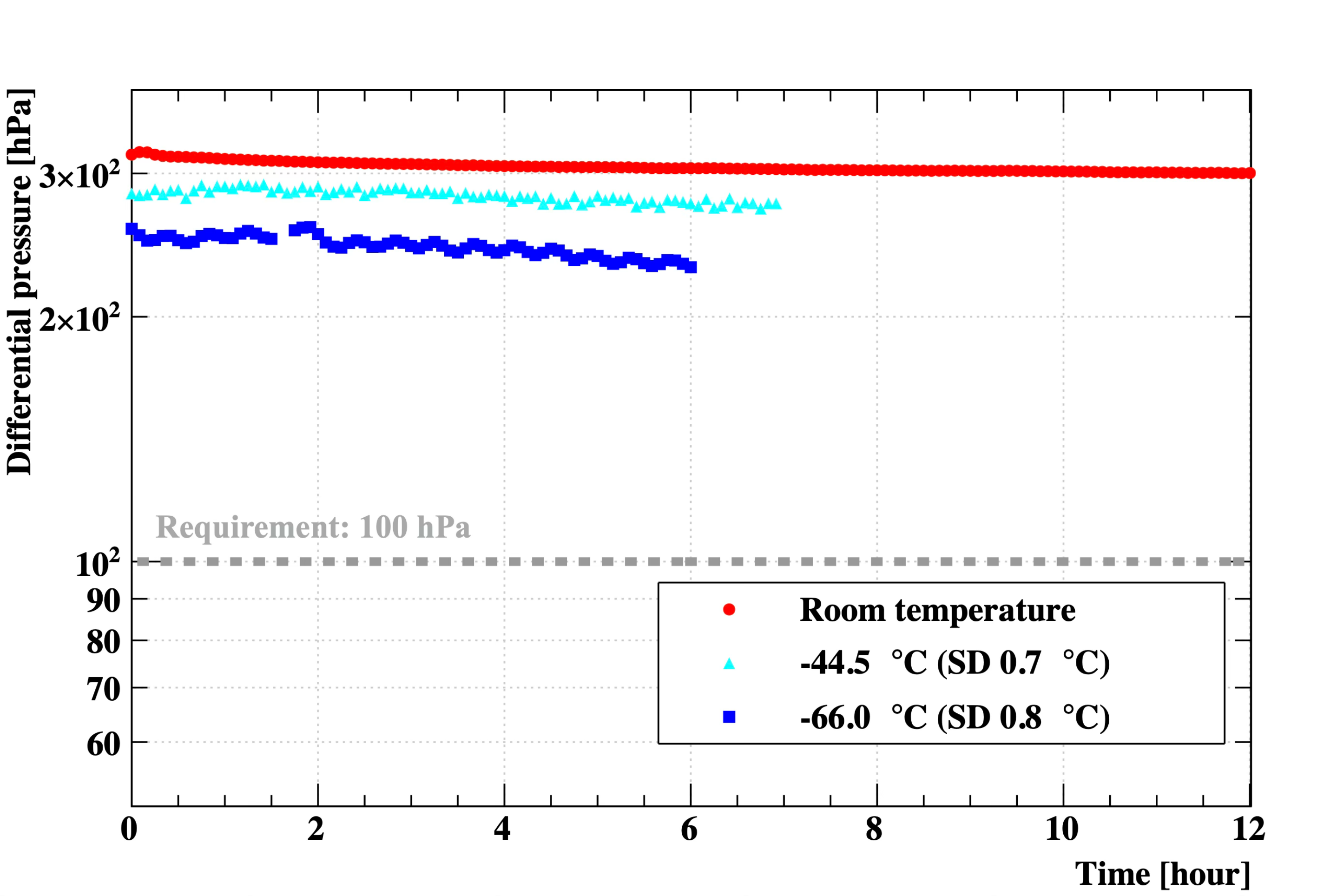}
% "\includegraphics" from the "graphicx" permits to crop (trim+clip)
% and rotate (angle) and image (and much more)
\caption{\label{fig:pressure-test} Pressure-retention performance of the GRAINE 2023 pressure-vessel gondola at room and low temperatures. The differential pressure between the inside and outside of the vessel is plotted as a function of time after pressurization. The red circles, cyan triangles, and blue squares indicate the results obtained at room temperature, a mean temperature of $-44.5^{\circ}$C with a standard deviation of $0.7 ^{\circ}$C, and a mean temperature of $-66.0^{\circ}$C with a standard deviation of $0.8^{\circ}$C, respectively. Even at a mean temperature of $-66.0^{\circ}$C, the differential pressure remained well above the requirement of 100 hPa for approximately 6 h, corresponding to the expected duration of the nighttime portion of the balloon flight.
}
\end{figure}

The differential pressure between the inside and outside of the vessel was measured at room temperature, at a mean temperature of $-44.5 ^{\circ}$C with a standard deviation of 0.7 $^{\circ}$C, and at a mean temperature of $-66.0 ^{\circ}$C with a standard deviation of 0.8 $^{\circ}$C. Figure \ref{fig:pressure-test} shows the differential pressure as a function of time after pressurization. Under the low-temperature conditions, measurements were continued for approximately 6 h. At both $-44.5 ^{\circ}$C and $-66.0 ^{\circ}$C, the differential pressure remained well above 100 hPa throughout the 6-h test period. The differential pressures after 6 h were approximately 96\%, 96\%, and 92\% of their initial values at room temperature, $-44.5 ^{\circ}$C, and $-66.0 ^{\circ}$C, respectively. Thus, even at a mean temperature of $-66.0 ^{\circ}$C, the complete pressure-vessel assembly, including the airtight membrane, membranous shell, heat-sealed joints, and sealing components, maintained the pressure-retention performance required for the expected nighttime flight duration.
\par
The differential-pressure relief valve was separately tested at $-60 ^{\circ}$C. The valve operated normally under the low-temperature condition and also provided the required pressure-retention performance when closed.
\par
These room- and low-temperature tests demonstrated that the GRAINE 2023 pressure-vessel assembly maintained a differential pressure above the ground-test criterion of 100 hPa for more than 40 h at room temperature and throughout the approximately 6-h low-temperature test at a mean temperature of $-66.0 ^{\circ}$C.
In addition, the differential-pressure relief valve was verified to provide both normal opening and closing operation and the required pressure-retention performance at $-60 ^{\circ}$C, confirming that the pressure-retention and pressure-control systems of the GRAINE 2023 pressure-vessel gondola retained the required functionality under low-temperature conditions.

\section{First overnight balloon flight of GRAINE and in-flight performance of the GRAINE 2023 pressure-vessel gondola}

GRAINE 2023 was the first overnight balloon experiment of the GRAINE project, continuing observations after sunset and until after sunrise on the following morning. This section describes the overview of the GRAINE 2023 balloon flight, the altitude and residual atmospheric pressure, the temperature environment experienced by the pressure-vessel gondola, and its in-flight pressure-retention performance. Based on these measurements, the design and ground-test results of the large-scale pressure-vessel gondola described in Section 2 are evaluated under the actual flight environment.

\subsection{Overview of the GRAINE 2023 balloon flight}

\begin{figure}[htbp]
\centering % \begin{center}/\end{center} takes some additional vertical space
\includegraphics[width=0.95\linewidth]{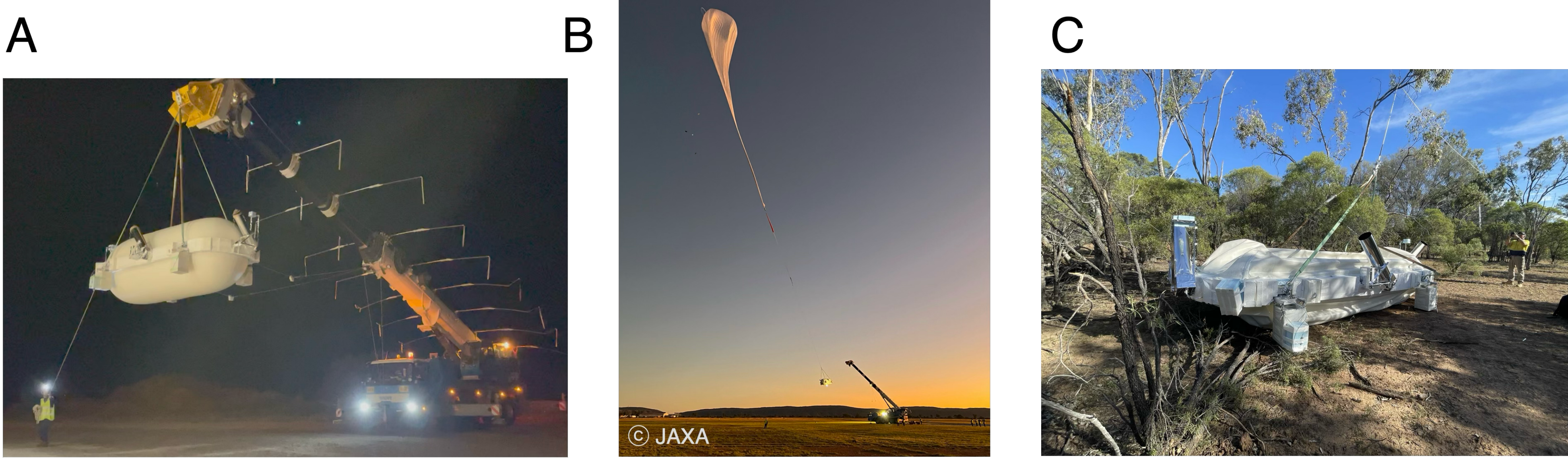}
% "\includegraphics" from the "graphicx" permits to crop (trim+clip)
% and rotate (angle) and image (and much more)
\caption{\label{fig:balloon_experiment} Pre-launch preparation, launch, and recovery of the GRAINE 2023 payload.
(A) The fully assembled payload during final pre-launch preparation at 00:50 local time on 30 April 2023.
(B) Launch from the Alice Springs Balloon Launching Station at 06:32 local time on 30 April 2023. Photograph courtesy of JAXA.
(C) The pressure-vessel gondola at the recovery site after landing near Longreach, Queensland.}
\end{figure}

The GRAINE 2023 balloon experiment was conducted at Alice Springs, Northern Territory, Australia, on April 30, 2023. Figure \ref{fig:balloon_experiment} shows the GRAINE 2023 payload during final pre-launch preparation, at launch, and after recovery.
\par
The GRAINE 2023 payload was launched at 06:32 on April 30, 2023, and reached a level-flight altitude of approximately 36 km about 2 h later. All times given in the text and figures in this section are in Australian Central Standard Time (ACST; UTC+9:30), corresponding to the local time at Alice Springs. The balloon subsequently continued its eastward level flight until payload cutdown at 08:47 on May 1. The total flight duration was approximately 27 h, including 24.3 h of level flight. This exceeded the total flight duration of 17.3 h achieved in GRAINE 2018 and represented the first GRAINE flight to continue from before sunset through the night until after sunrise on the following morning.

\begin{figure}[htbp]
\centering % \begin{center}/\end{center} takes some additional vertical space
\includegraphics[width=0.95\linewidth]{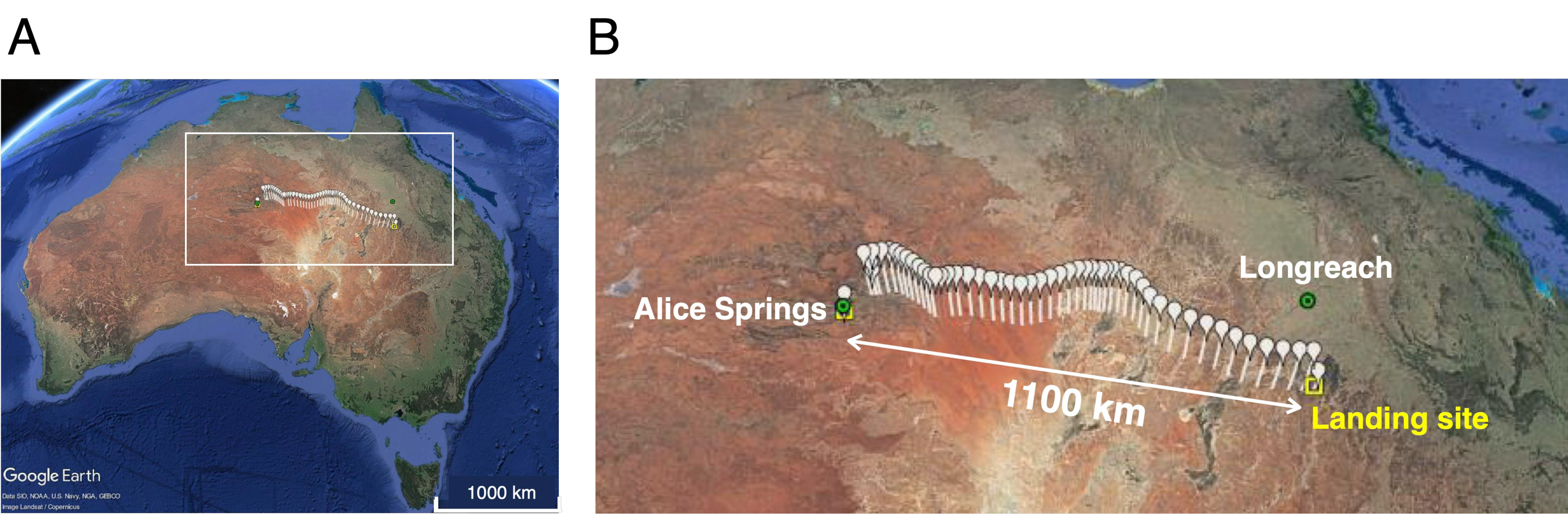}
% "\includegraphics" from the "graphicx" permits to crop (trim+clip)
% and rotate (angle) and image (and much more)
\caption{\label{fig:flight_path}Flight path of the GRAINE 2023 balloon experiment in Australia.
(A) Overview map of Australia showing the balloon flight trajectory. The white rectangle indicates the area enlarged in panel (B).
(B) Enlarged view of the flight trajectory from the launch site at Alice Springs to the landing site near Longreach. The straight-line distance between the launch and landing sites was approximately 1100 km.}
\end{figure}

Figure \ref{fig:flight_path} shows the flight path. The balloon traveled eastward from Alice Springs and landed at a location approximately 220 km south of Longreach, Queensland. The straight-line distance between the launch and landing sites was approximately 1100 km.
\par
During the flight, observations were conducted during time intervals in which the Vela pulsar and the Galactic center region entered the telescope field of view. Operation of the telescope system continued until approximately 08:00 on the following morning, providing observational data that included the nighttime visibility window of the Galactic center region.
\par
After flight termination, the payload descended by parachute, landed, and was recovered. The emulsion films carried inside the pressure vessel were transported to Japan under refrigerated conditions after recovery and were developed at Gifu University. The developed films were subsequently read out using automated track-readout systems at Nagoya University, and scientific analyses based on the reconstructed track data are in progress.

\subsection{Altitude and residual atmospheric pressure during the flight}

\begin{figure}[htbp]
\centering % \begin{center}/\end{center} takes some additional vertical space
\includegraphics[width=0.95\linewidth]{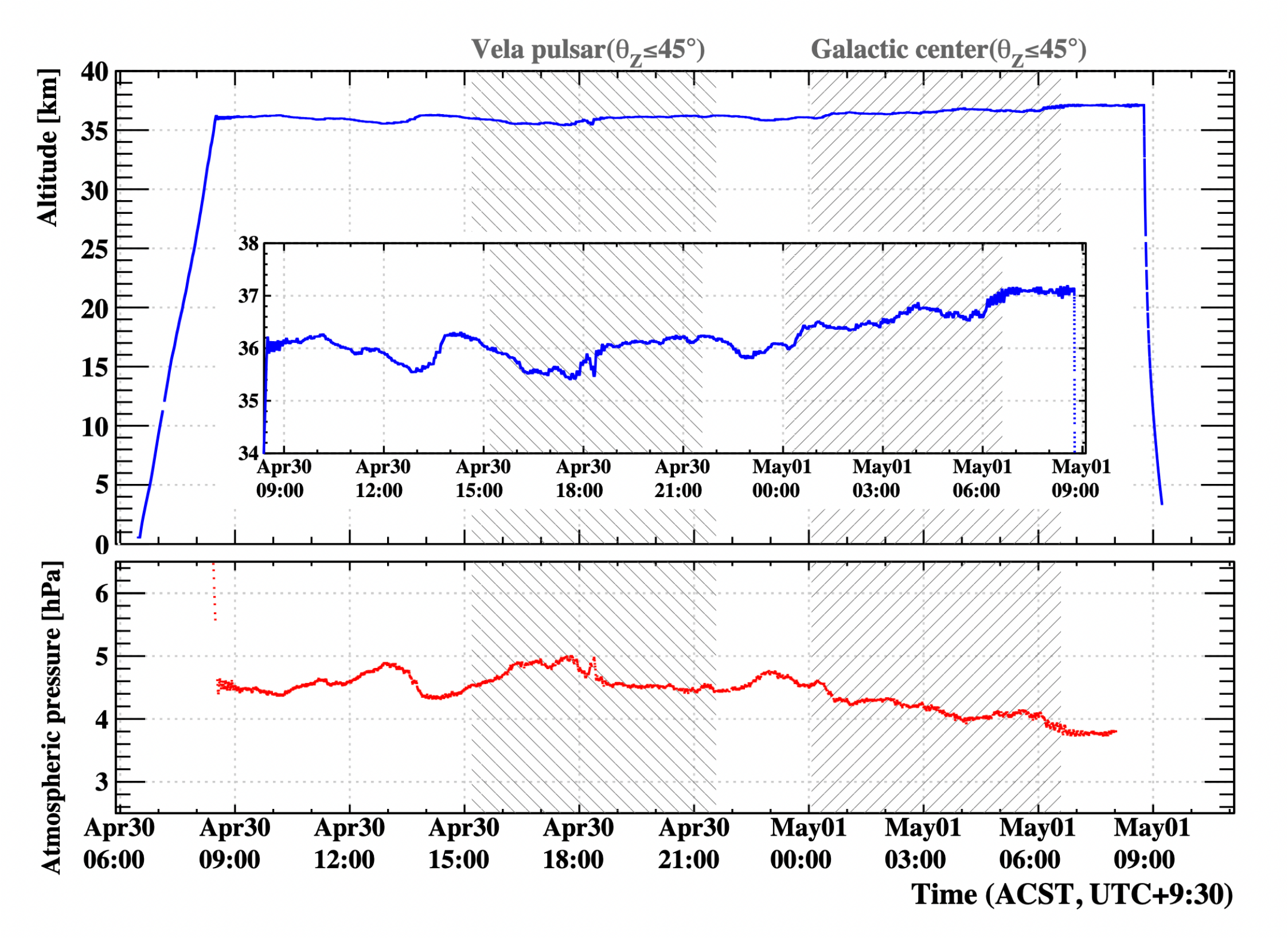}
% "\includegraphics" from the "graphicx" permits to crop (trim+clip)
% and rotate (angle) and image (and much more)
\caption{\label{fig:alt_press}Altitude and residual atmospheric pressure during the GRAINE 2023 balloon flight. The upper panel shows the altitude throughout the flight, and the inset provides an enlarged view of the level-flight period. The lower panel shows the residual atmospheric pressure at the payload altitude. The hatched regions indicate the time intervals during which the source zenith angle $\theta_z$ was less than or equal to $45^\circ$ for the Vela pulsar and the Galactic center. All times are given in Australian Central Standard Time (ACST, UTC+9:30).}
\end{figure}

Figure \ref{fig:alt_press} shows the altitude and residual atmospheric pressure during the GRAINE 2023 balloon flight. During level flight, the altitude varied between approximately 35.4 and 37.2 km. The upper panel of Figure \ref{fig:alt_press} shows the altitude throughout the flight, with an enlarged view of the level-flight period shown in the inset. The hatched regions indicate the time intervals during which the zenith angles of the Vela pulsar and the Galactic center were less than or equal to 45$^{\circ}$.
\par
The lower panel of Figure \ref{fig:alt_press} shows the ambient atmospheric pressure at the payload altitude. During level flight, the ambient pressure varied between 3.7 and 5.0 hPa. Observation of the Galactic center region therefore required not only a long-duration balloon flight but also stable operation of the telescope under the low-temperature and low-pressure nighttime environment, with the absolute pressure inside the pressure vessel maintained above 100 hPa.

\subsection{Temperature environment during the overnight flight}
\begin{figure}[htbp]
\centering % \begin{center}/\end{center} takes some additional vertical space 
\includegraphics[width=0.95\linewidth]{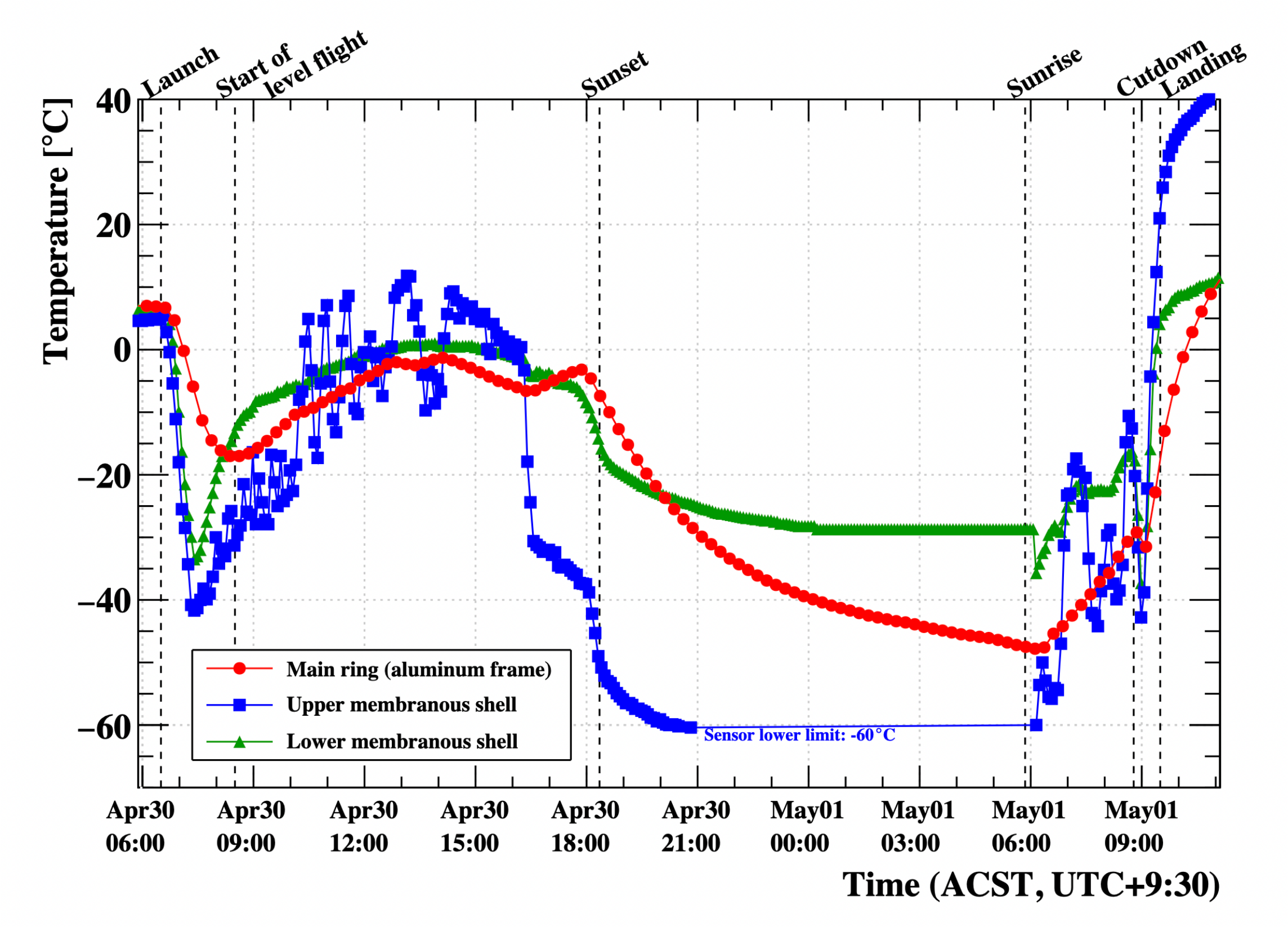}
% "\includegraphics" from the "graphicx" permits to crop (trim+clip)
% and rotate (angle) and image (and much more)
\caption{\label{fig:temperature}Temperatures of the GRAINE 2023 pressure-vessel gondola during the balloon flight. The red circles, blue squares, and green triangles show the temperatures measured on the main aluminum ring and the upper and lower membranous shells, respectively. The vertical dashed lines indicate the launch, start of level flight, sunset, sunrise, balloon cutdown, and landing. During the nighttime level-flight period, the upper-shell temperature reached the sensor lower limit of $-60^{\circ}$C. All times are given in Australian Central Standard Time (ACST, UTC+9:30).}
\end{figure}

Figure \ref{fig:temperature} shows the temperatures measured on the GRAINE 2023 pressure-vessel gondola during the flight. Temperatures were monitored at three locations: the aluminum-alloy main ring, the upper membranous shell, and the lower membranous shell.
\par
During ascent after launch, the temperatures of all three locations decreased as the ambient atmospheric temperature decreased. During daytime level flight, the temperatures increased and fluctuated because of solar irradiation. After sunset, the temperatures decreased again. In particular, the temperature of the upper membranous shell decreased rapidly and reached approximately $-60^{\circ}$C during the nighttime period. The lower membranous shell and main ring also showed gradual temperature decreases between sunset and sunrise.
\par
The temperature environment experienced during the flight was within the range evaluated in the ground tests: the full-scale low-temperature environmental test at a mean temperature of $-66.0^{\circ}$C described in Section 2.5 and the constant-load test of the heat-sealed membranous-shell joints at $-70^{\circ}$C described in Section 2.4. The ground low-temperature tests therefore covered the temperature conditions encountered during the actual overnight flight.

\subsection{In-flight pressure performance of the pressure-vessel gondola}
\begin{figure}[htbp]
\centering % \begin{center}/\end{center} takes some additional vertical space
\includegraphics[width=0.95\linewidth]{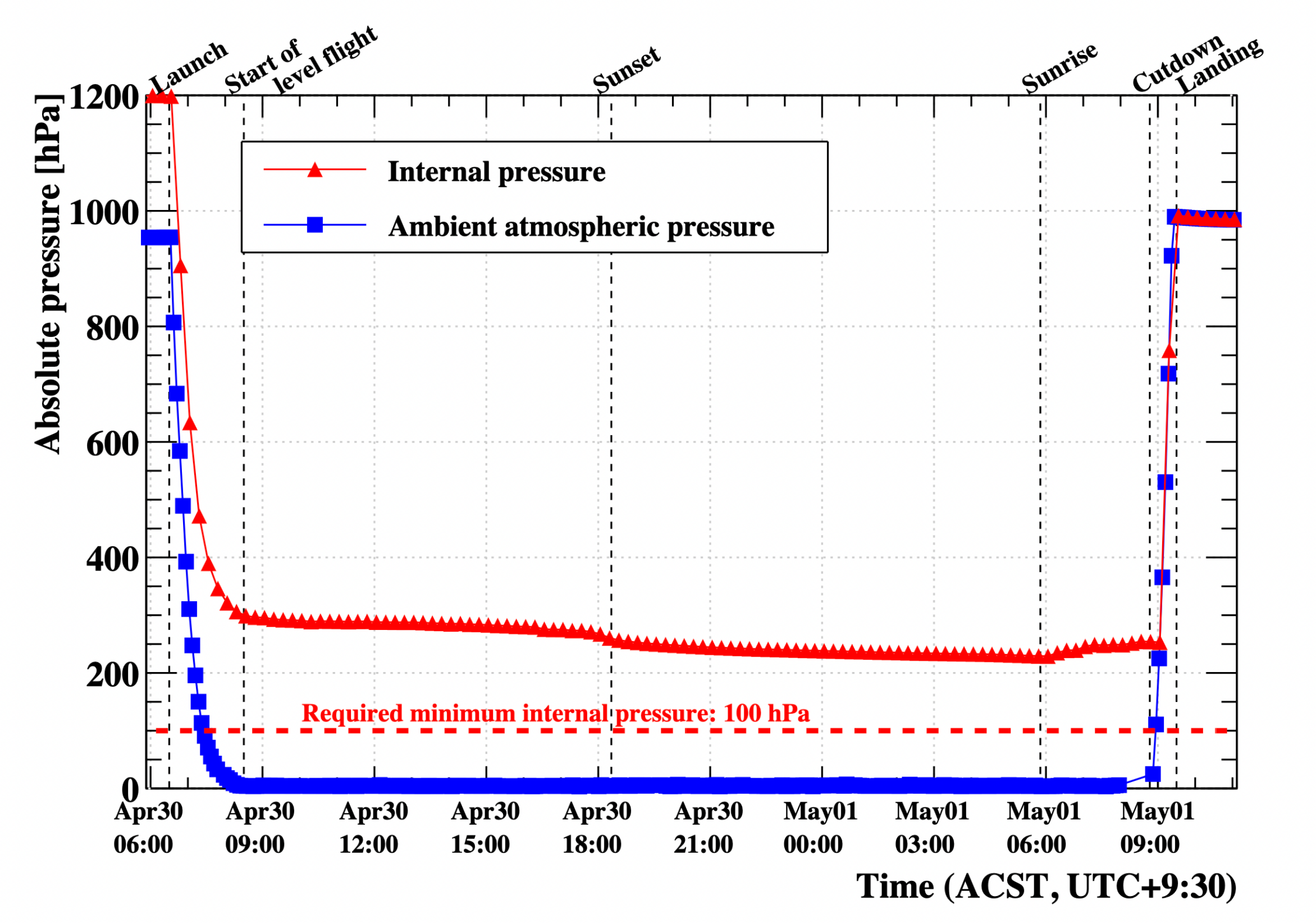}
% "\includegraphics" from the "graphicx" permits to crop (trim+clip)
% and rotate (angle) and image (and much more)
\caption{\label{fig:filight-pressure}Internal and ambient atmospheric pressures during the GRAINE 2023 balloon flight. The red triangles show the absolute pressure inside the pressure-vessel gondola, and the blue squares show the ambient atmospheric pressure outside the vessel. The horizontal red dashed line indicates the required minimum internal absolute pressure of 100 hPa. The internal pressure remained above this requirement throughout the level-flight period, including the nighttime portion of the flight. The vertical dashed lines indicate the launch, start of level flight, sunset, sunrise, balloon cutdown, and landing. All times are given in Australian Central Standard Time (ACST, UTC+9:30).
}
\end{figure}

Figure \ref{fig:filight-pressure} shows the absolute pressure inside the pressure-vessel gondola and the ambient atmospheric pressure measured during the GRAINE 2023 flight. Immediately after launch, the internal pressure was close to the atmospheric pressure at ground level. As the balloon ascended and the ambient pressure rapidly decreased, the differential pressure between the inside and outside of the vessel increased. When the differential pressure reached the operating pressure of the relief valve, the valve opened and released air from the vessel. Consequently, the internal pressure decreased with decreasing ambient pressure during ascent.
\par
After the start of level flight, the absolute pressure inside the vessel gradually decreased from approximately 300 hPa. The internal pressure was 260 hPa at sunset and 229 hPa at sunrise. The pressure decrease during the night corresponded to the decrease in vessel temperature shown in Figure \ref{fig:temperature}, and the internal pressure increased again after sunrise as the temperature rose. This behavior indicates that the in-flight pressure variation reflected not only gradual gas leakage but also changes in the temperature of the gas inside the vessel. The internal pressure remained well above the required minimum of 100 hPa.
\par
The absolute pressure inside the vessel remained above 100 hPa throughout the approximately 24.3-h level-flight period. No abrupt decrease in internal pressure was observed during the nighttime period, even when the upper membranous-shell temperature reached approximately $-60^{\circ}$C. These results demonstrated that the GRAINE 2023 pressure-vessel gondola, integrating the large-scale configuration, lightweight load-bearing structure, newly developed membranous-shell material, and low-temperature pressure-retention design described in Section 2, maintained the pressure environment required for operation of the 2.5-m$^{2}$ emulsion gamma-ray telescope throughout the first long-duration overnight flight of GRAINE.

\section{Conclusion}
GRAINE 2023 achieved the first overnight balloon flight of the GRAINE project using an emulsion gamma-ray telescope with a total aperture area expanded to 2.5 m$^{2}$, continuing observations through the night until after sunrise on the following morning. This flight enabled long-duration scientific observations with the large-area GRAINE telescope and provided the GRAINE 2023 dataset for full-scale analyses of astrophysical and atmospheric gamma rays.
\par
To enable these observations, the balloon-style pressure-vessel concept demonstrated in the previous study was scaled up to a scientific payload approximately 5 m in overall length. The ground-support and transport functions were transferred to a detachable transport cart to reduce the flight mass, and a new aluminum-alloy gondola ring structure was designed to support the two telescope units. The load-bearing structure was evaluated through finite-element analysis and full-scale tensile tests of the suspension points. In addition, SHL-300MDL was developed as a new membranous-shell material, reducing the areal density from 0.1 g/cm$^{2}$ for the previous SHL-300 to 0.067 g/cm$^{2}$, while satisfying the required membrane strength and maintaining the integrity of the heat-sealed joints. These design developments increased the telescope aperture area by a factor of 6.6, from 0.38 m$^{2}$ in GRAINE 2018 to 2.5 m$^{2}$ in GRAINE 2023, while limiting the increase in pressure-vessel gondola mass from 147 kg to 179 kg, corresponding to approximately 22\%.
\par
Long-duration pressure-retention tests at room temperature and full-scale low-temperature environmental tests were performed using the completed GRAINE 2023 pressure-vessel gondola. At room temperature, the differential pressure between the inside and outside of the vessel remained above 100 hPa for more than 40 h.
Even at a mean temperature of $-66.0 ^{\circ}$C, the vessel maintained a differential pressure above 100 hPa for approximately 6 h, demonstrating the low-temperature pressure-retention performance of the complete flight assembly, including the airtight membrane, membranous shell, heat-sealed joints, and sealing components.
\par
The GRAINE 2023 payload was launched from Alice Springs, Australia, on April 30, 2023, and achieved a total flight duration of approximately 27 h, including 24.3 h of level flight at an altitude of approximately 36 km. Although the temperature of the upper membranous shell reached approximately $-60 ^{\circ}$C during the night, the absolute pressure inside the pressure vessel remained above the operational requirement of 100 hPa throughout the level-flight period. The flight therefore demonstrated that the developed large and lightweight pressure-vessel gondola could provide the pressure environment required for stable operation of the 2.5-m$^{2}$ emulsion gamma-ray telescope under the low-temperature stratospheric conditions encountered during an overnight flight.
\par
After the flight, the emulsion films were recovered, developed, and scanned, and scientific analyses using the resulting track data are ongoing \cite{Yamamoto,Minami,Usuda}. 
In particular, a dedicated analysis of the Galactic center region using the GRAINE 2023 data is being prepared for publication.
Preliminary results from the GRAINE 2023 data on the Vela pulsar, and atmospheric gamma rays have been reported at international conferences \cite{NakamuraICRC,Nagahara}. These analyses mark the beginning of full-scale scientific observations with the GRAINE emulsion gamma-ray telescope expanded to a total aperture area of 2.5 m$^{2}$.
\par
The GRAINE project aims to further develop its scientific observations through repeated balloon experiments with large-area emulsion gamma-ray telescopes from different observation sites selected according to the target sources and observing conditions \cite{TakahashiASR,RokujoICRC}. 
The large and lightweight pressure-vessel gondola established in this work, together with its validation through ground tests and the overnight balloon flight, provides a common technical basis for future repeated observations.
By accumulating observational data over successive balloon flights, GRAINE will progressively expand its sub-GeV--GeV gamma-ray science.

%\appendix
\section*{Acknowledgment}
%\label{app1}
We would like to thank T. Kudo and R. Nishimura of the Equipment Development Support Section, Nagoya University, and T. Kawai of Japan Neutron Optics Inc. (J-NOP) for their valuable support in the design and evaluation of the gondola ring structure. We also thank H. Toyoda, F. Takeda, and T. Higashihata of Taiyo Kogyo Corporation for their contributions to the development of SHL-300MDL and the shells. We are grateful to the late K. Suzuki of Tokai Body for his invaluable support in providing access to and operating a rapid-cooling refrigerated container capable of reaching temperatures below $-60^{\circ}$C.
We thank C. Ikeda and the staff of the Scientific Ballooning Research and Operation Group of ISAS/JAXA for providing the scientific balloon (DAIKIKYU) flight opportunity and GPS data, and for their support during the balloon campaign.
This work was supported by JSPS KAKENHI Grant Numbers 17H06132, 18H05210, 23H00116, 21H04472, 20H01915, 24K00661, 22K20382, 22KJ2237, and 24KJ1273; the Nohmura Foundation for Membrane Structure's Technology; the DAIKO FOUNDATION; the FOUNDATION OF PUBLIC INTEREST OF TATEMATSU; the Nagoya University KMI Overseas Dispatch Program for Young Researchers; and the Nagoya University IMaSS Joint Research Program.

%%       \cite{<label>} ==> [1]

%%



\section*{Declaration of generative AI and AI-assisted  technologies in the manuscript preparation process}
During the preparation of this work, the authors used OpenAI ChatGPT for language editing to improve the clarity and readability of the manuscript. After using this tool, the authors reviewed and edited the content as needed and take full responsibility for the content of the published article.

\end{document}